# Surface resistance measurements of HTS thin films using SLAO dielectric resonator

Mohan V. Jacob[1], Janina E. Mazierska[1], Kenneth Leong[2], Dimitri O. Ledenyov[1] and Jerzy Krupka[3]

1) Electrical and Computer Engineering Department, School of Engineering, James Cook University, Townsville, QLD4811, Australia.
2) National Institute of Standards and Technology, Boulder, CO80305, U.S.A.
3) Instytut Mikroelektroniki i Optoelektroniki Politechniki Warszawskiej, Koszykowa 75, 00-662, Warszawa, Poland.

*Abstract*-surface resistance of HTS films is typically measured using Sapphire dielectric rod resonators enclosed in a copper cavity. In this paper we present surface resistance measurements of *YBa$_2$Cu$_3$O$_{7-\delta}$* films using Strontium Lanthanum Aluminate (*SLAO*) at a resonant frequency of *18.2 GHz*. We have performed the error analysis of the cavity loaded with *SLAO* dielectric rod and also verification measurements using two Sapphire (*Al$_2$O$_3$*) rod resonators operating at resonant frequencies of *24.6 GHz* and *10 GHz* respectively. Good agreement between the values of *Rs* of two sets of *YBa$_2$Cu$_3$O$_{7-\delta}$* films measured using the *SLAO* and the Sapphire dielectrics has been obtained after a frequency scaling of *Rs* was applied. Using different dielectric rods of the same size in the same cavity for measurements of *Rs* of *HTS* films it is feasible to do microwave characterization of the same films at differing frequencies.

## Introduction

The use of High Temperature Superconducting (*HTS*) planar filters in base station receivers increased significantly during the last few years e.g. [1]. The low insertion loss and sharp skirts are one of the most important characteristics of superconducting filters and their values depend on the surface resistance of the *HTS* thin films used. In order to fabricate *HTS* filters with very low insertion loss, it is important to identify *HTS* samples with low surface resistance at the frequency of interest. The surface resistance can be measured using different methods such as the cavity method [2], planar resonator method [3] and the dielectric resonator method [4]–[7]. The sensitivity and the error of the measurements are different for each technique [5]. The Sapphire dielectric resonator technique has been widely accepted for characterization of *HTS* samples due to its high sensitivity and the accuracy [5]. To characterize *HTS* thin films at different frequencies, different Sapphire rods and different cavities are required. However instead of using Sapphire rods of different diameters it is feasible to use different dielectric materials of the same dimensions in the same cavity.

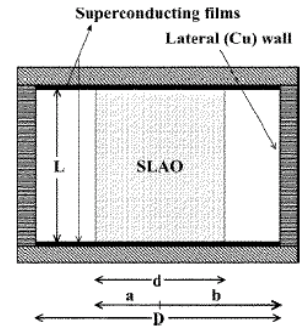

*Fig. 1. Schematic diagram of a SLAO dielectric resonator.*

In this paper we present results of microwave characterization of *YBa$_2$Cu$_3$O$_{7-\delta}$* films, using a *SrLaAlO$_4$* (*SLAO*) and a Sapphire (*AO*) rods of *3 mm* height and *5mm* diameters at frequencies of *18 GHz* and *24.6 GHz*.

## Dielectric resonator technique for microwave characterization of HTS films

The schematic diagram of the Hakki–Coleman (*H–C*) dielectric resonator used in this work is shown in Fig. 1. The resonator consisted of a *SLAO* dielectric rod of permittivity (of *16.8* between two *HTS* films enclosed in a copper cylindrical cavity. The copper cavity had the diameter of *9.5 mm* and a height of *3 mm*. Geometrical factors of the resonator with both dielectrics are given in Table I.

We have used the *TE* mode for our measurements. The resonant frequency of the *TE* mode resonator was calculated using the first root of the following transcendental equation [6]

$$k_{\rho 1}J_0(k_{\rho 1}b)F_1(b) + k_{\rho 2}J_1(k_{\rho 1}b)F_0(b) = 0 \quad (1)$$

where

$$F_0(\rho) = I_0(k_{\rho 2}\rho) + K_0(k_{\rho 2}\rho)\frac{I_1(k_{\rho 2}a)}{K_1(k_{\rho 2}a)}$$

$$F_1(\rho) = -I_1(k_{\rho 2}\rho) + K_1(k_{\rho 2}\rho)\frac{I_1(k_{\rho 2}a)}{K_1(k_{\rho 2}a)} \quad (2)$$

$$k_{\rho 1}^2 = \frac{\omega^2 \varepsilon_r}{c^2} - k_z^2, \quad k_{\rho 2}^2 = k_z^2 - \frac{\omega^2}{c^2}, \quad k_z = \frac{\pi}{L}$$

(2)

and $\omega$ is the angular frequency ($2\pi f$), $c$ is velocity of light, $\varepsilon_0$ is free space permeability, $\varepsilon_r$ is the real part of relative permittivity of the sample and $J_0$, $J_1$, $I_0$, $I_1$, $K_0$, $K_1$, denote corresponding *Bessel* and *Hankel* functions.

| Dielectric | $A_S$ | $A_M$ | $\rho_e$ |
|---|---|---|---|
| SrLaAlO$_4$ | 212.7 | 20241 | 0.98 |
| Al$_2$O$_3$ | 280.6 | 22319 | 0.97 |

*Tab. 1. Geometric factors and $\rho_e$ of the H-C resonator.*

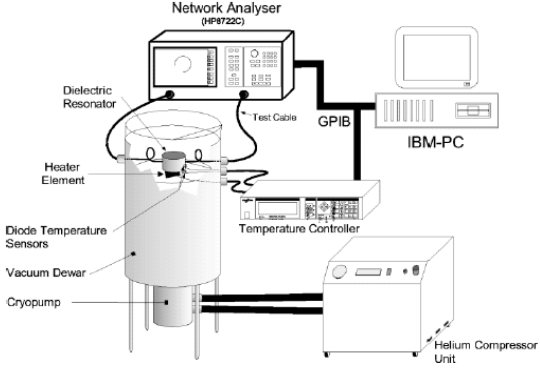

*Fig. 2. The experimental set-up to measure Rs of YBa$_2$Cu$_3$O$_{7-\delta}$ films with the SLAO resonator.*

The measurement system consisted of a Network Analyzer (*HP 8722C*), Temperature Controller (*Conductus LTC-10*), Vacuum Dewar, close cycle refrigerator (*APC-HC4*) and a computer as shown in Fig. 2.

The method of the surface resistance measurements was in principle the same as used by several groups e.g. [6], [7]. The main advantage (and difference) of our measurements was that the unloaded $Q_0$-factor was calculated from the exact equation, namely [8], [9]

$$Q_0 = Q_L(1 + \beta_1 + \beta_2) \quad (3)$$

where $\beta_1$ and $\beta_2$ are coupling coefficients of the resonator to the external circuitry. The loaded $Q_L$-factor and the coupling coefficients were obtained from multi-frequency measurements of $S_{21}$, $S_{11}$, and $S_{22}$ parameters measured around the resonance, using the Transmission Mode Q-Factor Technique [9], [10]. The *TMQF* method accounted for noise, delay due to uncompensated transmission lines and its frequency dependence, and crosstalk, which occurred in measurement data, and hence it provided accurate values of surface resistance.

The dielectric resonator loaded with the superconducting samples was mounted in the vacuum dewar and cooled down to *13 K*. The *S*-parameters were measured around the resonance from *13 K* to *85 K*. The $Q_0$-factor was calculated using the *TMQF* method for all temperatures.

The average surface resistance ($R_S$) of the *HTS* thin films was calculated using software *SUP12* [11], based on the well known equation [6]:

$$R_S = A_S \left[ \frac{1}{Q_0} - \frac{R_M}{A_M} - \rho_e \tan\delta \right] \quad (4)$$

where $R_M$ is the surface resistance of the metallic part of the cavity, $A_S$ and $A_M$ are the geometric factors of the superconducting part and metallic parts of the cavity and $\rho_e$ is the electric energy filling factor.

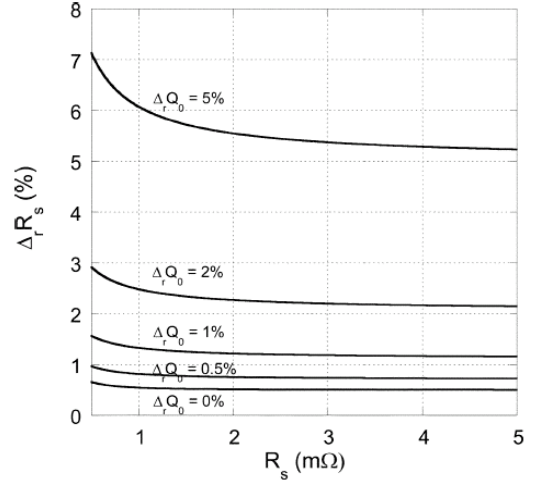

*Fig. 3. The most probable error in surface resistance of HTS measured with the SLAO rod for different uncertainties in $Q_0$ at 18.2 GHz.*

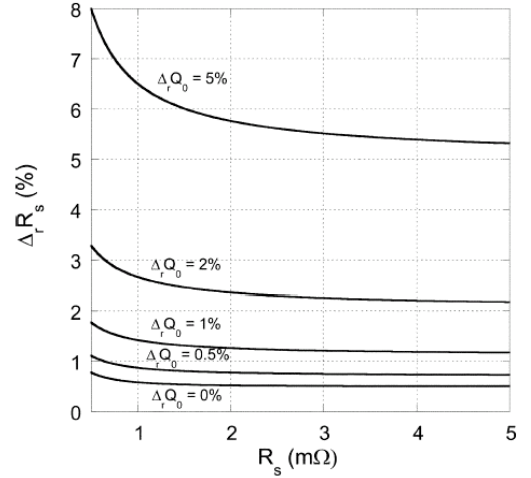

*Fig. 4. The most probable error in $R_S$ of A films measured with AO rod for varying uncertainties in $Q_0$ at 24.6GHz.*

Geometric factors $A_S$, $A_M$ and $\rho_e$ used in (4) were computed using the incremental frequency rules as [6] follows:

$$A_S = \frac{\omega^2 \mu_0 / 4}{\partial \omega / \partial L} \quad (5)$$

$$A_M = \frac{\omega^2 \mu_0 / 2}{\partial \omega / \partial a} \quad (6)$$

$$\rho_e = 2 \left| \frac{\partial \omega}{\partial \varepsilon_r} \right| \frac{\varepsilon_r}{\omega}. \quad (7)$$

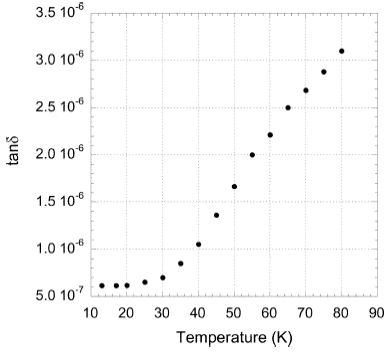

*Fig. 5. The loss tangent of the SLAO rod as a function of temperature.*

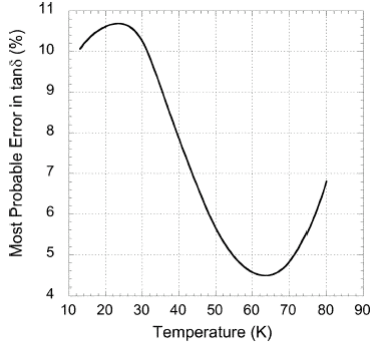

*Fig. 6. The most probable error in loss tangent of SLAO rod.*

### Error analysis of Rs measurements using the SLAO resonator

The most probable error in measured surface resistance using the dielectric resonator technique depends on surface resistance of the films and the cavity walls, loss tangent of the dielectric, geometrical factors of the resonator and uncertainties of their values. The error $\Delta R_S/R_S = \Delta_r R_S$ can be calculated on the basis of the following equation [5]:

$$\Delta_r R_S = \left[\left(1+\frac{R_M A_S}{R_S A_M}+\frac{A_S}{R_S Q_d}\right)^2 \left|\frac{\Delta Q_0}{Q_0}\right|^2 + \left|\frac{\Delta A_S}{A_S}\right|^2 + \left(\frac{A_S}{R_S Q_d}\right)^2 \left|\frac{\Delta Q_d}{Q_d}\right|^2 + \left(\frac{R_M A_S}{R_S A_M}\right)^2 \left(\left|\frac{\Delta R_M}{R_M}\right|^2 + \left|\frac{\Delta A_M}{A_M}\right|^2\right)\right]^{1/2}$$
(8)

where $Q_d = (\rho_d \tan\delta)^{-1}$; For the general error analysis of the *SLAO* and *AO* resonators we have assumed uncertainty in factor of *1%*, uncertainty in geometric factors of *0.5%*. The loss tangent of *SLAO* was assumed as $3\times10^{-6}$ and of *AO* as $1.25\times10^{-7}$ with the *6%* and *25%* uncertainty respectively. The surface resistance of copper was assumed to be *20 m$\Omega$* at *18.2 GHz* and *23.6m$\Omega$* at *24.6 GHz*. Figs. 3 and 4 show the most probable error in $R_S$ at *18.2 GHz* with the *SLAO* rod and at *24.6GHz* with the *AO* rod respectively.

In practice the loss tangent of the *SrLaAlO$_4$* rod varied from $6\times10^{-7}$ to $3.1\times10^{-6}$, and its uncertainty $\Delta_r \tan\delta$ was between *11.7%* and *4.5%*, when temperature changed from *13 K* to *80 K* as shown in Figs. 5 and 6.

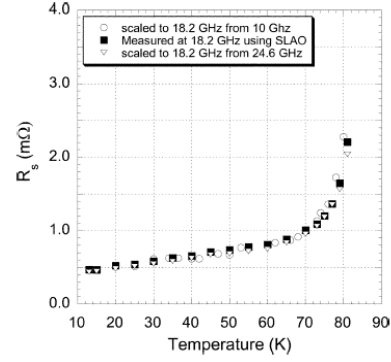

*Fig. 7. The surface resistance of YBa$_2$Cu$_3$O$_{7-\delta}$ films A at 18.2 GHz.*

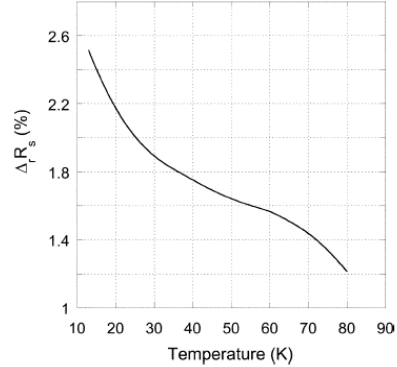

*Fig. 8. The $R_S$ error analysis with the SLAO resonator at 18.2 GHz.*

This was assessed on the basis of performed measurements of $\tan\delta$ as a function of the temperature, using *HTS* films with known surface resistance of *1.3%* uncertainty and the following equation [12]:

$$\Delta_r \tan\delta = \left[\left|\frac{\Delta\rho_e}{\rho_e}\right|^2 + \left(\left|\frac{-1}{Q_0\rho_e\tan\delta}\right|\left|\frac{\Delta Q_0}{Q_0}\right|\right)^2 + \left(\frac{R_{SS}}{A_S\rho_e\tan\delta}\left(\left|\frac{\Delta R_{SS}}{R_{SS}}\right|+\left|\frac{\Delta A_S}{A_S}\right|\right)\right)^2 + \left(\frac{R_{SM}}{A_M\rho_e\tan\delta}\left(\left|\frac{\Delta R_{SM}}{R_{SM}}\right|+\left|\frac{\Delta A_M}{A_M}\right|\right)\right)^2\right]^{1/2}.$$
(9)

### Results and discussion of Rs measurements of HTS films with SLAO and AO rods

Two sets (*A*: *1".* diameter and *B*: *0.5".* diameter) of thermally co-evaporated *YBa$_2$Cu$_3$O$_{7-\delta}$* films on *LaAlO$_3$* substrates were tested using the *SLAO* dielectric resonator. Measured surface resistance of the set *A* at frequency of *18.2 GHz* using the *SrLaAlO$_4$* dielectric rod is shown on Fig. 7 as squares. The $R_S$ of the same films measured with the *Sapphire* rod in the same cavity scaled to *18.2 GHz* from *24.6 GHz* is shown by triangles. The circles in the Fig. 7 represent $R_S$ of the same films measured at *10 GHz* with a different *Sapphire* rod (*12.32 mm* in diameter and *7.42 mm* in height) in a bigger cavity of *24.05 mm* in diameter scaled to *18.2 GHz*. The agreement in $R_S$ values measured with *SLAO* and *Sapphire* resonator is good. The deviation in $R_S$ was approximately *2%* for temperatures from *13 K* to *70 K* and approximately *8%* in the temperature range from *70K* to *80 K*.

The uncertainty analysis of $R_S$ of the films *A* using the *SLAO* rod and assuming the $\tan\delta$ and its uncertainty as in Fig. 5 and 6 gives the uncertainty in $R_S$ of films *A* presented in Fig. 8. The $\Delta_r R_S$ varies between *2.5%* at *13K* to *1.2%* at *80K* assuming *1%* uncertainty in the $Q_0$-factor.

On the basis of performed measurements of the films *A* with *SLAO* resonator at *18.2 GHz* and *AO* resonators at frequencies *10 GHz* and *24.6 GHz*, it is feasible to present the frequency dependence of the $YBa_2Cu_3O_{7-\delta}$ film (Fig. 9). We have obtained the frequency dependence of *2.1* ($f^{2.1}$) for $R_S$.

Fig. 10 shows the of $YBa_2Cu_3O_{7-\delta}$ films *B*, half inch diameter, measured at *18.2 GHz* frequency. The squares represent the measured values $R_S$ using the *SLAO* rod and the triangles represent the using the *AO* resonator at frequency of *24.5 GHz* and scaled to *18.2 GHz*. The deviation in $R_S$ values was approximately *3%* for temperatures from *13 K* to *70 K* and *10 %* in the temperature range from *70 K* to *80 K*.

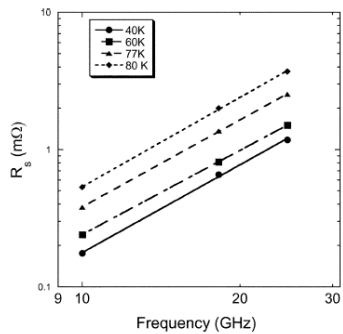

*Fig. 9. Frequency dependence of $YBa_2Cu_3O_{7-\delta}$ films A at temperatures 40 K, 60 K, 77 K and 80 K.*

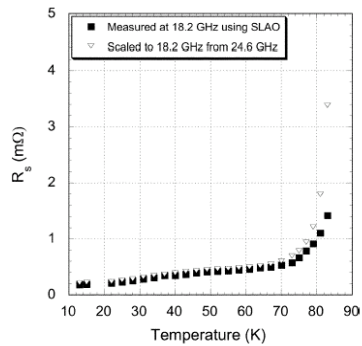

*Fig. 10. Surface resistance of $YBa_2Cu_3O_{7-\delta}$ films B at 18.2 GHz.*

## Conclusion

The surface resistance of two sets of $YBa_2Cu_3O_{7-\delta}$ samples was measured using the $SrLaAlO_4$ resonator at frequency of *18.2 GHz* and the $Al_2O_3$ resonator at frequencies of *10 GHz* and *24.6 GHz*. We have obtained two tests frequencies using same copper cavity and two dielectric materials of the same dimensions. The uncertainty analysis for the *18.2 GHz SLAO* resonator showed that the maximum most probable error in $R_S$ was *2.5%* (Fig. 8), while for the *24.6 GHz Sapphire* resonator the maximum $\Delta_r R_S$ was *1.8%* assuming $\Delta_r Q_0$ of *1%*. The verification measurements showed that difference between $R_S$ values measured with *SLAO* and *AO* was below 2% for temperatures up to *70 K* when the frequency scaling to the power 2.1 was used. Therefore by swapping the *Sapphire* rod with the *SLAO*, we will be able to measure the $R_S$ of $YBa_2Cu_3O_{7-\delta}$ thin films at two different frequencies without compromising the accuracy of measurements.